\documentclass[preprint]{revtex4-1} 
\usepackage{amsfonts} % needed for bold Greek, Fraktur, and blackboard bold
\usepackage{graphicx}
\usepackage{amsmath, amssymb}
\usepackage{slashed}
\usepackage{braket}
\usepackage[bottom]{footmisc}
\graphicspath{  {images/} }
\begin{document}

% Be sure to use the \title, \author, \affiliation, and \abstract macros
% to format your title page.  Don't use lower-level macros to  manually
% adjust the fonts and centering.

\title{A thermodynamic origin for the Cohen-Kaplan-Nelson bound}
% In a long title you can use \\ to force a line break at a certain location.

\author{Satish Ramakrishna}
\email{ramakrishna@physics.rutgers.edu}
%optional
%\altaffiliation[permanent address: ]{101 Main Street, 
%  Anytown, USA} % optional second address
% If there were a second author at the same address, we would put another 
% \author{} statement here.  Don't combine multiple authors in a single
%\author statement.
\affiliation{Department of Physics \& Astronomy, Rutgers, The State University of New Jersey, 136 Frelinghuysen Road
Piscataway, NJ 08854-8019}
% Please provide a full mailing address here.

% See the REVTeX documentation for more examples of author and affiliation lists.

\date{31-Oct-2021}

\begin{abstract}

The Cohen-Kaplan-Nelson bound is imposed on the grounds of logical consistency (with classical General Relativity) upon local quantum field theories. This paper puts the bound into the context of a thermodynamic principle applicable to a field with a particular equation of state in an expanding universe. This is achieved without overtly appealing to either a decreasing density of states or a minimum coupling requirement, though they might still be consistent with the results described. We do so by defining an appropriate Helmholtz free energy which when extremized relative to a key parameter (the Hubble radius $L$) provides a scaling formula for the entropy with the Hubble radius (an exponent $r$ used in the text). We deduce that the CKN bound is one possible solution to this extremization problem (with $r=\frac{3}{2}$), but there are others consistent with $r=2$.

The paper establishes that the holographic principle applied to cosmology is consistent with minimizing the free energy of the universe in the canonical ensemble, upon the assumption that the ultraviolet cutoff is a function of the causal horizon scale.

\end{abstract}
\maketitle

\section{INTRODUCTION}

The Cohen-Kaplan-Nelson \cite{CKN} bound hints at a connection between the ultraviolet cutoff applicable to local quantum field theory and the infrared cutoff applicable to the same theory. It is obtained by placing a limit on when gravitational effects should dominate the problem and collapse the relevant field configuration into a black hole. The limit is actually ensured by stating, by observation, that gravitational effects do not actually appear to play a significant role. The bound therfore constrains the behavior of local quantum field theories in the universe, by connecting the ultraviolet cutoff to the size of the causal horizon - the infrared limit of the theory. Essentially, the connection between the ultraviolet cutoff and the causal horizon size connects the number of possible internal states (which scales with some power of the length) to the smallest possible wavelength in the bulk of the region. 

Some recent work by Banks and Draper \cite{Banks} shows how to deduce the $\frac{3}{2}$-power of the entropy/length-scale dependence (that the CKN bound leads to) from an  appropriately chosen ansatz for the density of states in the bulk. 

%Some other work \cite{David} indicates how the bound might lead to a minimum coupling requirement akin to the weak gravity conjecture.

A parallel piece of (prior) work, especially that of Fischler and Susskind \cite{FischSuss} deduces various cosmological consequences of equations of state. It finds that the quadratic entropy/length-scale dependence required by holography is realized only with the equation of state $P=\rho$, which is achieved in the special case of a massless scalar field with no potential function, i.e., with only a kinetic term. They deduce this from a very general starting point using classical gravity, i.e., the Friedmann equation combined with the equation of state.

The question arises - how are the above pieces of work connected? Is it an accident that the CKN bound gives a $\frac{3}{2}$ power scaling with length for entropy while the Fischler-Susskind bound gives quadratic scaling for $w=1$ and $\frac{3}{2}$-power length scaling for $w=\frac{1}{3}$? Can one derive both from one framework? Are they actually just two ways of looking at the same problem?

\subsection{Summary of the work presented}

This work finds a common ground between the above sequence of thoughts as well as posits the CKN bound as the consequence of a physical principle. To this end, we frame the problem as one of  equilibrium thermodynamics of the entire universe bathed in a thermal gas that obeys, in the simplest case, the equation of state $P=w \rho$ for various values of $w$.  We define the Helmholtz free energy for this thermal gas and also (subsequently) apply the same condition to a radiation-dominated universe. From just this, we deduce that the dependence of entropy on the length scale is indeed the $\frac{3}{2}$ power. The resulting equation turns out to be {\it exactly} the CKN bound. 

We conclude that the CKN bound is exactly the equivalent of extremizing the free-energy in an expanding universe (exactly what it is extremized versus is explained later).  After this, we apply the methodology to the case $w=1$ and re-derive the Fischler/Susskind conditions (and the area-entropy connection), in the canonical ensemble, by minimizing the Helmholtz free energy. We also confirm the bound that they found in this case.

In what follows, we will focus on a parameter $r$, which is the power of the Hubble radius that the entropy scales in accordance with. Standard holographic ideas would require $r=2$. The CKN bound yields $r=\frac{3}{2}$. 
% ADDITION
As we will note in the discussions that follow, there are other solutions to the extremization problem that yield $r=2$ under differing assumptions about densities of states and non-linear equations of state.
% ADDITION

\section{The Fischler-Susskind bound}
Let us study the thermodynamics of an entropy-saturating field in an expanding universe as in \cite{FischSuss}. The authors impose the holographic condition that, operating with Planck units
\begin{eqnarray}
\sigma R_H^{D} < [a(t) R_H]^{D-1}
\end{eqnarray}
where $t$ is the cosmic time, $R_H= \int_0^t \frac{dt}{a(t)}$ is the horizon, $\sigma$ is the (assumed) constant comoving entropy density and $D$ is the number of spatial dimensions.
Saturating the limit is achieved by setting $a(t) \sim t^p \rightarrow p = \frac{1}{D}$, which is consistent with an entropy-saturating  equation of state $(P = w \rho$ with $w=1)$. In this limit, the R.H.S of Equation (1) is proportional to the square of $L=a(t) R_H$; in $3$ spatial dimensions, this is the area.

We can derive these from applying the thermodynamic principle of extremizing the free energy as follows. We consider the following cases, for the radiation gas ($w=\frac{1}{3}$) and the scalar field with only kinetic term ($w=1$). 

We will use the following general formulas, {\it {viz.}}
if $P= w \rho$, then, in $D$ spatial dimensions,  with $t$ the cosmic time and $a(t)$ the scale factor
\begin{eqnarray}
\rho \sim \frac{1}{a^{D (1+w)}}  \: \: \: , \: \: \: a(t) \sim t^{\frac{2}{D(1+w)}}  \: \: \: , \: \: \: T \sim  \frac{1}{a(t)^{D w}}
\end{eqnarray}

The causal horizon is defined as 
\begin{eqnarray}
R_H =  \int_0^t \frac{dt}{a(t)} \sim t^{1 - \frac{2}{D(1+w)}} \nonumber \\
L = a(t) \times R_H = t  \: \: \: \: 
\end{eqnarray}

In these terms, for a general equation of state (the temperature dependence is derived under the constant entropy density assumption \cite{Stornaiolo} ),
\begin{eqnarray}
\rho \sim \frac{1}{L^2}  \: \: \: , \: \: \: a(t) \sim L^{\frac{2}{D(1+w)}} \: \: \: , \: \: \: T \sim \frac{1}{L^{\frac{2 w}{1+w}}} \: \: \: , \: \: \: \rho \sim T^{\frac{1+w}{w}}
\end{eqnarray}

% ADDITION
These formulas apply under the condition that the equation of state is linear, i.e., that $w$ is a number and not itself a function of $\rho$ or $a$. In addition, we are assuming homogeneity and isotropic expansion with constant entropy density; hence, the quantities $\rho, a(t), T$ cannot deviate from the functional form above. We are exploring in a future publication the effect of non-linear equations of state upon the arguments in this paper (some remarks are made in Sections IV and VI).
% ADDITION

In what follows, we will keep dimensions in check with appropriate powers of $M_P$, the Planck energy and work in natural units ($c=1, \hbar=1)$.

\subsection{The thermodynamic framework}

We apply the usual laws of thermodynamics to an expanding universe bathed in a thermal gas with equation of state $P=w \rho$ and a temperature $T$. We work in the canonical ensemble - this consistent choice implies we are assuming that we can define an equilibriated temperature for this thermal gas, though we allow processes (eg. expansion) where the temperature changes. We define the Helmholtz free energy
\begin{eqnarray}
F= \rho V - T S
\end{eqnarray}
where $\rho$ is the energy density, $V=L^3$ is the volume of the causal horizon, $T$ is the temperature of the field and $S$ its entropy.

%  ADDITION

All these quantities (including the entropy $S$) are functions of $L$, the size of the causal horizon. We are asserting that the universe remains in equilibrium while it is expanding. In equilibrium, the free energy is at an extremum and the universe's expansion needs to be consistent with the free energy being extremized. All the quantities in the free-energy expression are functions of $L$. It is natural that we should extremize the free energy with respect to the causal horizon $L$.

% ADDITION

These relationships are detailed in Equation (4) above for a linear equation of state. For instance, the temperature in a radiation-dominated universe, is $T \sim\frac{M_P}{ a(t)} \sim \sqrt{\frac{M_P}{L}}$. In addition, $S$, the entropy, is assumed to depend on it as $S \sim L^r$.  Since entropy is dimensionless, we will write it as $S = (M_P L)^r$.

The energy density has two components, generically. There will be a temperature dependent piece and there will be a vacuum energy piece. For instance, in a radiation-dominated universe with an ultraviolet cutoff $\Lambda$, the total energy is 
\begin{eqnarray}
\rho V = L^3 \bigg( \sigma T^4 +  \Lambda^4 \bigg)
\end{eqnarray}
where $\sigma$ is the Stefan-Boltzmann constant.  It is tempting to think that this term is dominated in the late(r) universe by the ultraviolet cutoff term, but we are going to allow $\Lambda$ to scale with some power of $L$, say $\Lambda(L) \sim M_P (\frac{1}{M_P L})^{\mu}$. Here we have inserted $M_P$ to produce proper dimensions for $\Lambda$.

% REMOVED
%We require that the universe stays close to equilibrium while expanding, hence we should be able to establish a condition upon the free energy - it should be extremized with %respect to $L$.
%REMOVED

% ADDITION
As explained earlier, we will extremize the free energy with respect to $L$ - physically, this implies that the universe is in equilibrium as it expands.
% ADDITION
Since $\rho V$ and $TS$ are proportional to powers of $L$, the extremization produces, upto factors of order $\sim 1$, that $\rho V \sim T S$.

% ADDITION
Under this analysis, the vacuum energy is a function of $L$, so we are forced to assume if will participate and equilibriate with the fluid that pervades the universe. It is indeed an energy density and in this picture, it does participate in the equilibrium thermodynamics of the universe.
% ADDITION

Hence, we are going to impose the uniform condition $\rho V \sim T S$.

\section{The case $w=\frac{1}{3}$}

We are going to make some simplifications to match dimensions correctly. We start with (as explained previously),
\begin{eqnarray}
\rho V = TS  \: \: \: \: \: \rightarrow  \: \: \: \: \:   \: \: \: \: \:  \: \: \: \: \:  L^3 \bigg( \sigma T^4 +  \Lambda^4(L) \bigg) \sim \sqrt{\frac{M_P}{L}} \:  (M_P L)^r
\end{eqnarray}
This approximate equality can be written as 
\begin{eqnarray}
\sigma \bigg(\frac{T}{M_P}\bigg)^4 + \bigg(\frac{\Lambda(L)}{M_P}\bigg)^4 \sim \frac{1}{(M_P L)^{\frac{7}{2}-r}}
\end{eqnarray}
We notice that
\begin{eqnarray}
\bigg(\frac{T}{M_P}\bigg)^4 \sim  \frac{1}{(M_P L)^2}    \: \: \: \: \:  ,   \: \: \: \: \:  \bigg(\frac{\Lambda(L)}{M_P}\bigg)^4 \sim \bigg(\frac{1}{M_P L}\bigg)^{4 \mu}
\end{eqnarray}
By inspection, an easy solution to this equation by comparing powers of $L$ is given by $\mu = \frac{1}{2}$ and $r = \frac{3}{2}$. 

This case (with $r=\frac{3}{2}$) is also the one obtained by the CKN condition, i.e., since the two terms in Equation (8) have the same dependence on $L$, we can write (upto factors of order $\sim 1$),
\begin{eqnarray}
\bigg(\frac{\Lambda(L)}{M_P}\bigg)^4 \sim \frac{1}{(M_P L)^{2}}
\end{eqnarray}
which is {\it exactly} the CKN condition.

% ADDITION

Of course, this makes sense - in this particular case, with $w=\frac{1}{3}$, the causal horizon is $L$. In addition, the energy inside the universe is $\sim \rho L^3 \sim \frac{1}{L^2} L^3 \sim L$. The Schwarzschild radius is of the same order as the causal horizon, which is indeed the CKN condition, so it is indeed not surprising that this case leads to the above result.

% ADDITION

As it turns out (and this is originally due to the model's authors\cite{CKN}), this solution can also be obtained by counting states in a self-consistent manner. If we count the states in the quantum field theory, we'd obtain (from a standard local field theory calculation), and approximate the solution for $\Lambda$ by ignoring the non-vacuum energy part,
\begin{eqnarray}
(M_P L)^r = (dimensionless \: constants) \: \times \: L^3 \int_{\frac{1}{L}}^{\Lambda(L)} dk \: k^2 \sim L^3 \Lambda^3(L) = (M_P L)^3 (\frac{\Lambda(L)}{M_P})^3 \nonumber \\
\rightarrow  (M_P L)^r \sim  (M_P L)^3 \frac{1}{(M_P L)^{\frac{3}{4} (\frac{7}{2}-r)}}   \: \: \: \: \:   \: \: \: \: \:   \: \: \: \: \:   \: \: \: \: \:  \nonumber \\
\rightarrow r = \frac{3}{2}  \: \: \: \: \:   \: \: \: \: \:   \: \: \: \: \:   \: \: \: \: \:  \: \: \: \: \:   \: \: \: \: \:   \: \: \: \: \:   \: \: \: \: \:  \: \: \: \: \:   \: \: \: \: \:   \: \: \: \: \:   \: \: \: \: \: 
\end{eqnarray}
which again indicates that the entropy scales as the $3/2$ power of the Hubble length scale, as is known. Note that we have derived this from an application of the minimization of free energy principle rather than the CKN condition, though, as may be noted.

\section{Other solutions to Equation (8)}
Apart from the condition $\mu = \frac{1}{2}$, we can consider other solutions, classified as 
\begin{enumerate}
\item $\mu > \frac{1}{2}$: Here, in Equation (8), the temperature term dominates for large $L$, we can easily see that $r=\frac{3}{2}$ is the only consistent solution.
% CORRECTION
\item $\mu < \frac{1}{2}$: In this case, the ultraviolet cutoff term dominates for large $L$. We might then proceed to find a relation between $\mu$ and $r$, however, note that by Equation (4), the energy density is forced to scale as $\rho \sim \frac{1}{L^2}$. This constrains the problem massively and prevents there being any other solution  (other than $r=\frac{3}{2}$) in the linear equation of state case.
\item If we assumed that the density of states (as used in Equation (11)) is $\sim k^m$ ($m=2$ is the usual case) and we further write $\Lambda(L) \sim M_P \frac{1}{(M_P L)^q}$, we obtain the following consequences, i.e.,
\begin{eqnarray}
q (m+2) = 2 \: \: \: \: , \: \: \: \: 3 -q (m+1) = r \nonumber \: \: \: \: \: \: \:  \: \: \:  \: \: \:  \: \: \:  \: \: \:  \\
\rightarrow \: \: \: \:q = \frac{2}{m+2} \: \: \:  \: \: \:  \: \: \:  r = \frac{m+4}{m+2}  \: \: \: \: \: \: \:  \: \: \: \: \: \: \:  \: \: \: \: \: \: \:  \: \: \: \: \: \: \: 
\end{eqnarray}
which are obtained by requiring that the energy density scales as $\sim \frac{1}{L^2}$ while the number of states (the entropy) scales as $L^r$.
We then obtain different possible combinations of parameters (see Table 1). For consistency with the minimization of free energy principle, we still expect $r=\frac{3}{2}$. However, with a non-linear equation of state, we could get a different temperature dependence with $L$ (the Hubble horizon) and this can lead to other values of $r$ corresponding to different values of $m$ and $q$. This rich array of solutions is being explored\cite{SatishFut}, but a quick summary follows.

\subsection{Nonlinear equations of state}

 Generally speaking, with an equation of state, i.e.,
\begin{eqnarray}
P=(w_0+w_1 \rho) \rho \nonumber \\
T \sim \frac{1}{a^{D(w_0+w_1 \times a^{-D(1+w_0)})}}
\end{eqnarray}
hence, if we define $T \sim \frac{1}{L^{\gamma}}$ and if $w_1>0$, $\gamma > \frac{2 w_0}{1+ w_0}$ (conversely, $\gamma$ is smaller, if $w_1$ were negative). Then, we find 
\begin{eqnarray}
q = \frac{2}{m+2} \: \: \:  \: \: \:  \: \: \:  r = \frac{m+4}{m+2}  \: \: \: \: \: \: \:  \: \: \: \: \: \: \:  \: \: \: \: \: \: \:  \: \: \: \: \: \: \: \nonumber \\
1+\gamma = r  \: \: \: \: \: \: \:  \: \: \: \: \: \: \:  \: \: \: \: \: \: \:  \: \: \: \: \: \: \:  \: \: \: \: \: \: \: \: \:  \: \: \: \: \: \: \: \: \: \nonumber \\
\rightarrow q = \gamma  \: \: \: \: \: \: \:  \: \: \: \: \: \: \: \: \:  \: \: \: \: \: \: \:  \: \: \: \: \: \: \: \: \:  \: \: \: \: \: \: \:  \: \: \: \: \: \: \: \: \: 
\end{eqnarray}
Interestingly, the inverse power of temperature is the same as the inverse Hubble-length scaling exponent of the ultraviolet cutoff. In general, we'd expect $r$ to be higher with an equation of state $P=(w_0+w_1 \rho) \rho$, if $w_1>0$.

\begin{table}[h!]
  \begin{center}
    \caption{Combinations of Density of States and Length Scale Dependence of $\Lambda$}
    \label{tab:table1}
    \begin{tabular}{l|c|r} % <-- Alignments: 1st column left, 2nd middle and 3rd right, with vertical lines in between
      \textbf{ (Density of States exponent)} & \textbf{Scaling of $\Lambda$} & \textbf{Entropy Scaling}\\
      \: \: \: \: \: \: \: \: \: \: \: \: \: $m$ &  $q$ &  $r$ \: \: \: \: \: \:  \\
      \hline
       \: \: \: \: \: \: \: \: \: \: \: \: \: 2 & $\frac{1}{2}$ & $\frac{3}{2}$\: \: \: \: \: \: \\
       \: \: \: \: \: \: \: \: \: \: \: \: \: 1 & $\frac{2}{3}$ & $\frac{5}{3}$\: \: \: \: \: \: \\
       \: \: \: \: \: \: \: \: \: \: \: \: \: 0 & 1 & 2\: \: \: \: \: \: \\
    \end{tabular}
  \end{center}
\end{table}
\end{enumerate}
% CORRECTION
To demonstrate the generality of the free-energy principle, we next apply it to the case of the ``perfect'' gas, i.e.,  with $w=1$, exemplified by the scalar field with only a kinetic term.

\section{The case $w=1$}

In this case, the total energy again is written as the sum of the field energy plus the vacuum energy upto the ultraviolet cutoff, i.e., 
\begin{eqnarray}
\rho V =  L^3 (\alpha T^2+\Lambda^4(L))
\end{eqnarray}
Again, the ultraviolet cutoff is assumed to possess some dependence upon $L$, of a form similar to the previous section, i.e., $\Lambda(L) \sim  M_P (\frac{1}{M_P L})^{\mu}$. 

The $T^2$ dependence in Equation (15) is a straightforward consequence of the thermodynamics of expansion, see Equation (4).
The equation has an appropriate (dimensionful) constant $\alpha$.  In this case, however, the temperature term in the universe dominated by this type of radiation is $T\sim\frac{M_P}{ a(t)^3} \sim \frac{1}{L}$. Again, assuming that the entropy $S \sim (M_P L)^r$, we write the equation in a dimensionally simple way (${\hat \alpha}$ is a dimensionless constant constructed from $\alpha$ and $M_P$), i.e., 
\begin{eqnarray}
 \big({\hat \alpha} \big(\frac{T}{M_P}\big)^2 + \big(\frac{\Lambda}{M_P}\big)^4\big) \sim   \frac{1}{(M_P L)^{4-r}}
\end{eqnarray}

A similar observation to the above is in order, i.e., 
\begin{eqnarray}
\bigg(\frac{T}{M_P}\bigg)^2 \sim  \frac{1}{(M_P L)^2}    \: \: \: \: \:  ,   \: \: \: \: \:  \bigg(\frac{\Lambda(L)}{M_P}\bigg)^4 \sim \bigg(\frac{1}{M_P L}\bigg)^{4 \mu}
\end{eqnarray}
Again, by inspection of the terms, one possible self-consistent solution is
\begin{eqnarray}
\mu = \frac{1}{2} \: \: \: \: \: , \: \: \: \: \:  r =2  
\end{eqnarray}
is a self-consistent solution. We thus find, in agreement with Fischler and Susskind \cite{FischSuss}, that the entropy/area relation is reached for the case $P=\rho$.

\section{Other solutions to Equation (16)}

Apart from the condition $\mu = \frac{1}{2}, r=2$, we can consider other solutions, classified as 
\begin{enumerate}
\item $\mu > \frac{1}{2}$: Here, in Equation (16), the temperature term dominates for large $L$, we can easily see that $r=2$ is the only consistent solution.
% CORRECTION
\item $\mu < \frac{1}{2}$: In this case, the ultraviolet cutoff term dominates for large $L$. We again find no other possible solutions (other than $r=2$) due to the constraints on the evolution of the energy density ($\rho \sim \frac{1}{L^2}$). Once again, the general analysis in the previous section applies and we can get other values of the entropy exponent $r$, with suitable choices of $q$ and $m$ and a density of states calculation. However, we still need to stay consistent with the minimization of free energy principle that we have mandated and we again find that with a non-linear equation of state, a modified dependence of the temperature $T$ upon $L$  and other values of the entropy exponent $r$.
\end{enumerate}
% CORRECTION

All this can be generalized to $D$ spatial dimensions, i.e., with the simplest solution in all cases,
\begin{eqnarray}
r =D -  \frac{2}{1+w}
\end{eqnarray}

In the terminology of Banks \& Draper \cite{Banks}, the power dependence of the density of states is derived from $L(\epsilon) \sim \frac{1}{\epsilon^n}$ and we will find $n = 1 + \frac{2}{3}(D -  \frac{2}{1+w})$.

\begin{figure}[h!]
\caption{Entropy Exponent vs. Pressure/Energy Density Ratio}
\centering
\includegraphics[scale=.5]{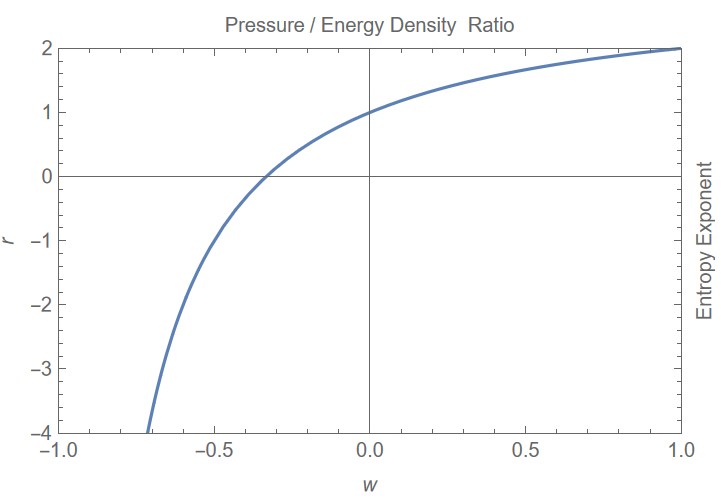}
\end{figure}

\section{Numerical Estimates}

The usual objections to the CKN bound are due to the implied size of the ultraviolet cutoff. If $\Lambda \sim \frac{1}{\sqrt{L}}$, then for the current Hubble radius, we would expect $\Lambda \sim 0.7 eV$. This is a regime that appears to be well tested to not be a sensible ultraviolet limit. However, if $\gamma = q = \frac{1}{3}$, for instance, which is a result consistent with a nonlinear equation of state for $w_1<0$, then we could well find $\Lambda \sim 1 \: GeV$, which is a more reasonable limit.

%With the result obtained after Equation (12), $\Lambda \sim \frac{1}{L^{3/8}}$, which leads to $\Lambda \sim 1 \: MeV$, which is closer to an  acceptable limit.

\section{Conclusions}

We have established a simple free-energy minimization principle in the canonical ensemble to derive the CKN bound. To some extent, these calculations use a blend of classical gravitational physics with standard quantum field theory. They show that (in this case, at least) applying the holographic bound to the problem is exactly the same as minimizing the free-energy of the universe in the canonical ensemble. In this manner, they are similar in spirit to the entropic views of gravity and immediately shed light on entropy-length scale relations. We extend the method to compute the bound applicable to other equations of state and demonstrate that one can deduce (in the radiation-dominated case) the entropy/length-scale relations without amending the density of states of the underlying quantum field theory. Adding the vacuum energy into the mix of energies in the thermodynamic problem allows one to enrich the space of possible entropy/length-scale relationships in a way that has not been previously appreciated.

\section{Data Disclosure}

Data sharing is not applicable to this article as no new data were created or analyzed in this study. No conflicts exist.

\section{Acknowledgments}

Detailed discussions with Tom Banks and Scott Thomas were especially useful. An especially lucid colloquium by Patrick Draper motivated this study and discussions with him after are also gratefully acknowledged. The hospitality of the Rutgers Physics Department and the NHETC is also acknowledged. A rather detailed and deep review by an anonymous reviewer is acknowledged for fixing an error in a preliminary version of this paper.

\end{document}